# Direct matter disassembly via electron beam control: Electron-beam-mediated catalytic etching of graphene by nanoparticles

*Ondrej Dyck,[1*] David Lingerfelt,[1] Songkil Kim,[2] Stephen Jesse,[1] and Sergei V. Kalinin,[1]*

[1] Center for Nanophase Materials Sciences, Oak Ridge National Laboratory, Oak Ridge, TN

[2] School of Mechanical Engineering, Pusan National University, Busan 46241, South Korea

*E-mail: dyckoe@ornl.gov

**Abstract**

We report electron-beam activated motion of a catalytic nanoparticle along a graphene step edge and associated etching of the edge. This approach enables beam-controlled etching of matter through activated electrocatalytic processes. The applications of electron-beam control as a paradigm for molecular-scale robotics are discussed.

**Introduction**

Since the emergence of nanoscience as an integrated field in late 1980s, the creation of nanometer-scale robotics capable of manipulating matter on the molecular- and atomic-scales has become one of the primary goals of nanotechnology, which has sparked new areas of research in molecular machines, nanoparticles, and self-assembly across the scientific landscape.[1,2] Such ideas have firmly captured the imagination of the general community as evidenced by multiple popular SciFI writers (e.g. Neal Stephenson's *The Diamond Age* and Michael Crichton's *Prey*). Nanoassemblers and disassemblers, "grey goo," medicinal nanorobots and related concepts have become a part of everyday vocabulary.

However, the expectations have been severely compromised by reality. Top-down fabrication of micro- and nano-electromechanical systems (NEMS and MEMS) encounter progressively growing complications, both on the fabrication and operation sides. Bottom-up assembly using biological motors and molecular machines is presently limited by synthesis and integration issues, with little progress achieved since the early work by Montemagno.[3,4] More importantly, even if the immediate synthetic and assembly bottlenecks are surpassed, the question of programming and operation of nanomachines remains an open issue and pose multiple questions. For example: Will such machines be capable of individual operation, possibly in communication with others? Will they be required to form and operate as a collective, swarm-like system? Will these be electrically powered machines or chemically powered, electrically controlled machines? In other words, what will be the source of energy (chemical, electrical) and control (external, internal, collective)?

In considering the basic principles and operation of nano- and molecular-scale robotics, we note that the key elements that define the operation of any active entity must include (a) a source of power, (b) locomotion, (c) sensors, (d) communication methods, and (e) signal processing and control (thinking). For signal processing and control, the technologies available at this point are the most advanced as a result of the incessant development of semiconductor-based electronics from the late 1940s until now. While integration of Si-based electronics with non-Si elements represents a complex problem, the technological basis for these developments is available and the scaling-down of the Internet of Things (IoT) technologies may provide a pathway for the emergence of customizable, flexible, small-scale control units.

(Loco)motion represents a second and significantly more complex issue. The solutions based on NEMS and MEMS generally scale poorly below the sub-micron level and are generally incompatible with operation in liquid and other non-vacuum environments. Biological systems including kinesine can, in principle, be incorporated in microfabricated assemblies; however, progress in this area has been slow. Eric Drexler popularized molecular machines and proposed the controlled chemical synthesis of molecular entities capable of autonomous motion and performing work, alternatively referred to as the nanomachines and nanoassemblers.[5] This concept immediately captured the imagination of a broad sector of the scientific community.[6,7] Experimentally, efforts towards light,[8,9] chemically,[10] and electrochemically[11,12] activated molecular machines were undertaken as recognized by Jean-Pierre Sauvage, Sir J. Fraser Stoddart, and Bernard L. Feringa's 2016 Nobel Prize for chemistry.

However, realization of this approach to molecular scale robotics requires the simultaneous solution of three problems, namely design of molecular blocks carrying required functionalities, development of the synthetic pathway, and assembly in operational supramolecular structures. The power of modern computational methods makes the former viable, if not yet fully accomplished. Synthesis represents a more complex problem, traditionally requiring outstanding organic synthesis intuition and a broad knowledge base. The recent establishment of reaction databases combined with advances in graphical search algorithms has enabled automatic identification of synthetic pathways for all known and many unknown (e.g., via retrosynthesis) compounds.[13,14] However, it is the probing and assembly of molecular machines into operational structures that remain a central issue.

We note that an interesting approach for enabling small-scale robotics is based on the combination of native biological systems (e.g., insects) and imposed cybernetic control, which are likely to result in viable technologies at the 100-micrometer range and above once the semiconductor control and integration issues are addressed.

Sensors and communication represent the next level of complexity. The presence of multiple chemical sensing technologies suggests that between chemical, optical, and electromagnetic channels, multiple solutions are available. The interesting corollary of communication is the fact that the "thinking" functionality does not have to be confined within an individual unit – rather it can be separated between multiple interacting units, giving rise to the collective intelligence swarm-like behavior adopted by many insect, bird, and fish species. Finally, an energy source is a necessary requirement for any mobile system. The source can be internal (battery, fuel cell) or external (light, RF field, availability of chemical species). Similarly, the power source can be electrical, chemical, light, or field driven.

Note that the biggest problem in the development of viable nano- and molecular-level robotics lies in the difficulties for integrating multiple dissimilar functionalities into progressively smaller volumes. Even in biological systems refined over billions of years of evolution, a reduction in size below millimeters typically leads to the obviation of higher control functions and transition to swarm behavior; at smaller length scales, the operational system becomes essentially hard-wired in the structure and the resulting units have extremely narrow (and hence difficult to control) functionalities, as exemplified by fages and viruses.

As an interesting development, we further note that in many areas there is a transition between localized behavior to that controlled and observed externally. Examples of light-driven and light controlled or magnetic field driven and controlled molecular machines exist.[15-17]

We propose that molecular molecular machines can be based on direct control with electron-beam (e-beam) probes. In this approach, the e-beam serves as (a) an energy source enabling machine operation, (b) a signal providing control, and (c), switching between control and imaging modes providing a read-out of machine action. Here, we demonstrate that the elementary stage of this process – electron beam induced motion of an atomic scale object – can be realized experimentally. This development builds upon recent work for e-beam applications for direct atomic fabrication.[18-24]

**Results and Discussion**

Here, we demonstrate and leverage the ability to switch on a catalytic etching process to drive the motion of a W nanoparticle attached to a graphene step edge using an electron beam (e-beam). Switching between lower dose imaging and concentrated targeting of the nanoparticle with the e-beam serves to transition between observation of the present state of the system and activation of the nanoparticle movement. Previously, several investigations on thermally activated nanoparticle etching of graphene through catalytic hydrogenation have shown that this technique can be used to control precise graphene edges, demonstrating high selectivity towards terminations. However, a byproduct of the etching process results in movement of the catalytic particles in the etching direction. Some nanoparticle materials that have been used to induce etching of graphane or graphene are Ni,[25,26] Fe,[27] Ag,[28] Co,[29] Pt, Ru, and PtRu.[30] In each of these cases the parameter controlling the etching process (e.g., sample temperature) was applied macroscopically (often *ex situ*), which affects all the nanoparticles equally. In the context of developing nanorobotic devices, a macroscopically applied process activation leads to significant challenges in terms of controllability.

Initiating a controllable process one particle (or unit) at a time and switching it on and off as desired is a first critical step on the pathway toward a nanorobotic device. Previous investigations into driving such a process have included x-rays, magnetic fields, light, acoustic waves, electric fields and thermal energy.[31-33] Here, we employ a strategy where the sample temperature is held below the activation energy for the catalytic hydrogenation process which causes nanoparticle movement, and explore whether the process can be activated by directed illumination from an incident focused e-beam in a scanning transmission electron microscope (STEM). In this configuration, specific individual nanoparticles can be targeted and the etching process, resulting in nanoparticle movement, can be monitored *in situ*.

Graphene films were transferred onto a Protochips Fusion heater chip via a liquid transfer method to control the sample temperature *in situ*. Transferred graphene samples are typically dirty after a liquid transfer;[34] thus, to remove most of the amorphous contaminant material and introduce multilayer graphene nano-islands the temperature was ramped *in situ* to 1200 °C using the heater chip and held constant for the duration of the experiment. An overview of this sample configuration is shown in Figure 1. The graphene film is suspended over a hole in the heater chip in Figure 1a. Multi-layer graphene nano-islands are adhered to the surface after heating to 1200 °C, likely held in place by defects in the main graphene sheet and a line/box profile (blue box in Figure 1b) illustrates the discrete steps in HAADF-STEM image intensity that is consistent with multilayer graphene. Bright nanoparticles (NPs) are also observed decorating the nano-islands. Electron energy loss spectroscopy (EELS) was performed on the NPs. In the energy range 0-2000 eV the only core loss peaks were from tungsten (Figure 1c,d) and carbon. From this information we infer that the nanoparticles are either tungsten or tungsten carbide. Measuring the lattice spacing from the image shown in Figure 2k and similar others (not shown) reveals a {110} plane spacing of 2.07 Å with a standard deviation of 0.07 Å. This is consistent (within 17 pm) with the expected plane spacing of body centered cubic tungsten metal (2.23 Å) observed along the [100] direction and with the expected crystal symmetry. This indicates that the particles are tungsten metal.

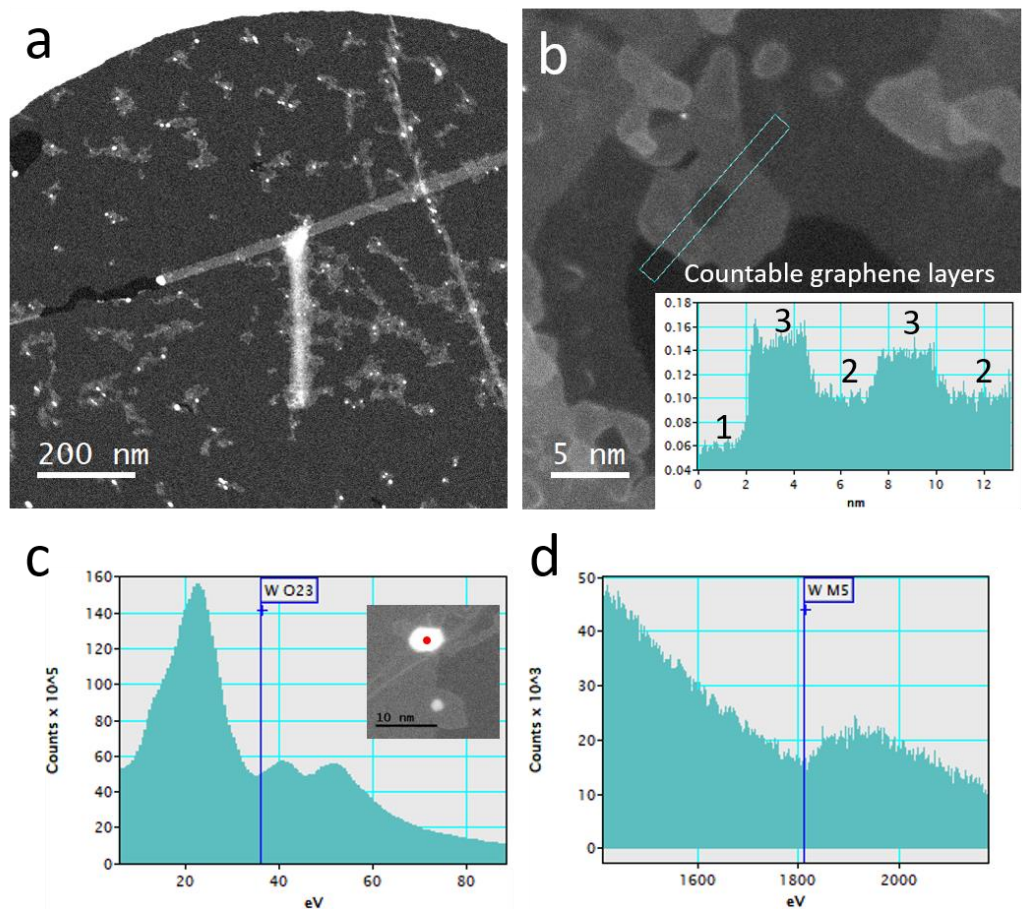

**Figure 1 Overview of the sample upon heating to 1200 ℃.** a) HAADF-STEM image of suspended graphene with residual contamination on the surface. b) HAADF-STEM image of contaminated region showing intensity steps indicative of multi-layer graphene. Inset is an intensity profile across the boxed region with the number of graphene layers labeled. c) EELS for Tungsten $O_{23}$ and d) Tungsten $M_5$ edges acquired with beam positioned on bright nanoparticle (red dot) shown inset in c), indicating nanoparticles decorating graphene are tungsten.

To observe e-beam effects on the NP-graphene interface, a sub-scan box was defined within the larger field of view (FOV) so the e-beam could be directed manually while a continuously updated sub-image was observed to track any sample changes. The details of this process are shown in Figure 2. The initial configuration, with the NP attached to several graphene nano-islands, is shown in Figure 2a. A few frames of the sub-image generated by the sub-scan are shown in Figure 2j-l. We observed that under the influence of e-beam irradiation, the NP begins to move out of the sub-scan region. The sub-scan area was manually repositioned to ensure continuous irradiation of the same NP and periodically, additional images were captured to document the sample state. First, we observed (Figure 2a-b) that the NP etched away the nano-island indicated by the dotted outline. Likewise, the transition from Figure 2b-c and c-d, the nano-islands outlined were etched away. This etching process caused the NP to move along the edge of the graphene and attach to another graphene island (Figure 2d-e). The NP etched its way along the graphene step edge to encounter another nanoplatelet Figure 2e, which was also etched away. A second example of this etching process is shown in Figure 2f-i.

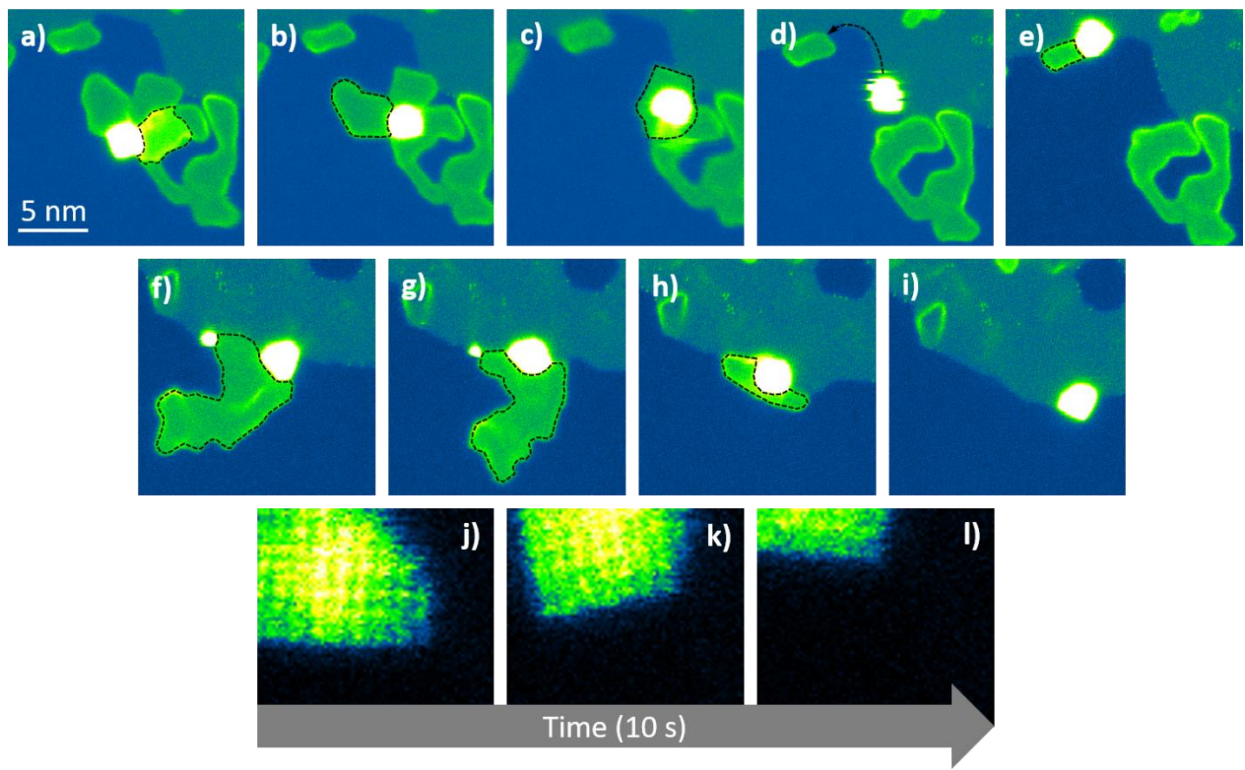

**Figure 2 e-beam-induced NP-graphene interface evolution.** a) HAADF-STEM image of initial configuration showing W-rich NP (white) surrounded by graphene nano-islands supported on main graphene film. Sub-scan box defined within larger FOV and manually positioned on NP. e-beam irradiation activated an etching process where graphene nano-islands are sequentially consumed. Outlined areas in each image indicate graphene nano-islands that are gone by next frame. b)-d) NP etched away graphene nano-island it was initially attached to and then attached itself to edge of larger sheet. In d)-e) the NP moved along graphene edge and attached to another nano-island. f)-i) NP etched away larger nano-island. j)-l) Sequence of sub-scan images acquired over NP showing its movement over 10 s relative to scan area.

A NP attached to a step edge away from any nanoplatelets is shown in Figure 3a. The sub-scan region was again positioned over the NP and as shown in Figure 3b-h, the NP etches a ~5 nm wide channel through the larger graphene bilayer island until finally stopping upon encountering a thicker multi-layer graphene area.

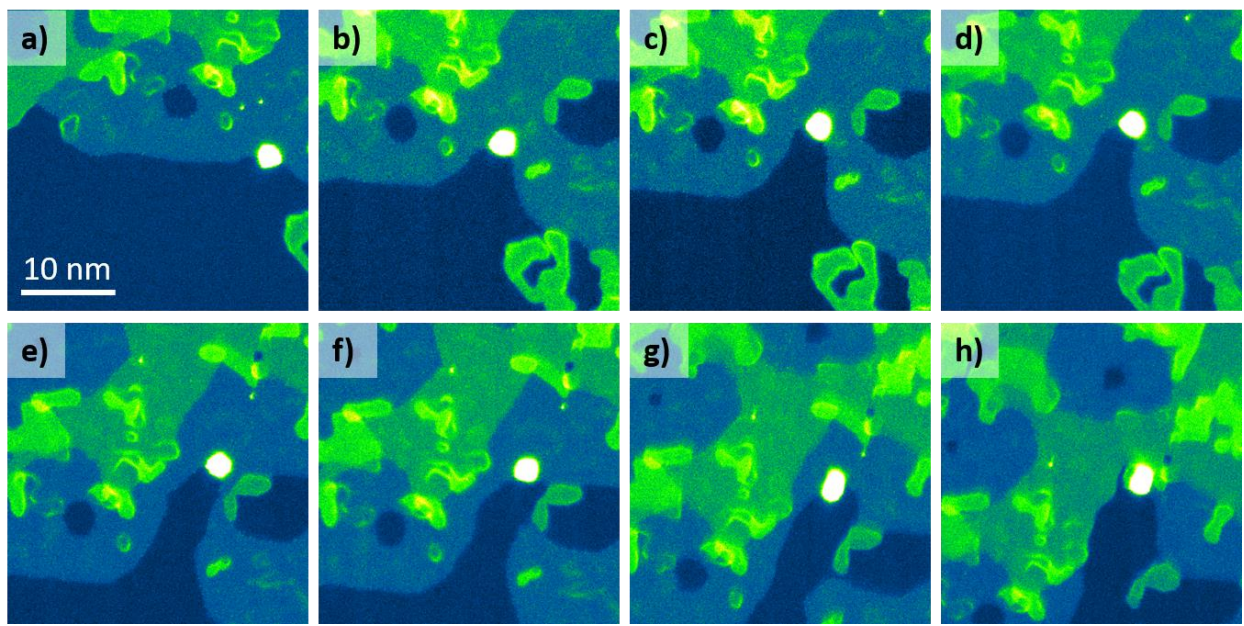

**Figure 3 W-rich NP etching a channel through a single layer of bilayer graphene.** a) Initial configuration shows NP attached to concave step/edge between single layer (darker blue – bottom) and bilayer graphene (lighter blue). b)-h) Irradiating NP by manually positioning sub-scan region caused NP to etch a ~5 nm channel through one layer of bilayer graphene. Etching rate slowed appreciably when NP contacted multilayer graphene island and experiment was stopped.

We note that the etching process observed in this case is associated with the disappearance of a single layer of the graphene without the formation of visible NPs or solid reaction products within the FOV or appreciable changes in the NP size. We hence argue that the process may involve the interaction with the residual hydrogen, and the formation of volatile hydrocarbons, as has been reported for other similar systems.[25-30] The etching process is activated by localization of the e-beam on the NP. If we assume that the etching process results from catalytic reduction of the graphene, employing hydrogen contaminants in the system as feedstock, small volatile hydrocarbons will be produced. The mechanism underlying this reaction is a familiar one in the study of heterogenous catalysis. Molecular hydrogen has long been known to add dissociatively to clean W surfaces[35] and the capacity for W to promote both hydrogenation[36] and hydrogenolysis[37] of hydrocarbons is well established.

Since etching can be promoted under thermal equilibrium at elevated temperatures, one must conclude that a non-thermal distribution of excited state populations induced through inelastic scattering cannot be the absolute requirement for promoting etching with electron beams; however, focusing an electron beam on a W-rich NP will only create a small degree of local heating that is unlikely to drive the catalytic hydrogenation.[38,39] Therefore, the population of high energy electronic/phonon levels excited in response to e-beam irradiation must open the system to reaction pathways that are thermally inaccessible.

Ideally one should be able to create a model of the system by examining the electronic/phonon response of a NP to the e-beam and the energetic pathways for graphene decomposition at a step/edge. Unfortunately, a full *ab initio* calculation of heterogenous catalysis at a metal surface remains a grand challenge in the physical sciences.[40] Affordable mean-field mixed quantum-classical simulation methods all fail categorically for metals decorated with small chemisorbed molecules.[41] For the present case, an explicit treatment of the correlated electronic and vibrational dynamics is required. Mixed quantum-classical methods that capture these correlations have been combined with simplified treatments of the excited-state electronic structure of a metal to simulate the electronic excitations induced by scattering of incident gas-phase molecules from a metal surface,[42] but methods capable of simultaneously treating the dynamic electronic and vibrational correlations and the "strong" (static) electronic correlation that emerges as covalent bonds are broken, have yet to be put forth.

While a first principles understanding of the catalytic activity of a metal surface is currently unavailable, some hints to the nature of the chemistry are provided by molecule-surface scattering experiments. The strong nonadiabatic coupling between short-lived excitons (electron-hole pairs) in the metal and the vibrational degrees of freedom of the adsorbate has been demonstrated through ultrafast vibrational relaxation of NO upon collision with gold[43] and through the "chemicurrents" induced upon adsorption of various small molecules onto silver.[44]

Another potential mechanism contributing to graphene etching by metal NPs at elevated temperatures (one that has yet to be explored in the literature) is that the NPs are not promoting heterogenous catalysis under the electron beam but are instead alloying with the carbon to form metal carbides. Carbonization of W surfaces is achievable by heating W treated with ethene (at room temperature) to 1000 K.[45] However, this mechanism is expected to reach a saturation point beyond which the uptake of carbon ceases and should result in an increase in nanoparticle size and, eventually, crystal structure. This was not observed in our experiments.

We note that these preliminary studies open a large parameter space for further exploration. For example, the reaction process could be significantly accelerated and more control could be attained through the introduction of a hydrogen gas source as opposed to relying on residual hydrogen within the STEM column. Similarly of interest is the role of the e-beam fluence and energy, studies that can be facilitated by rapid beam energy switching.

Overall, we demonstrate that the atomically focused electron beam of a STEM can be used to induce catalytic etching and nanoparticle movement, inducing local reactions and etching of the graphene step/edge or along single layers of multi-layer graphene. This beam induced motion and etching satisfies several of the conditions of the nanoscale robotics, namely the external power and certain level of control. Previously, such phenomena were explored in the context of light activated micromotors and robotics. Here, they are translated to the atomic level.

Furthermore, these studies suggest that in a system with multiple particles and edges their activity can be turned on and off individually, allowing for complex dynamic phenomena. These studies complement recent studies on e-beam-induced atomic motion and positioning and may offer a complementary paradigm for e-beam controlled manipulation and assembly.

**Acknowledgements**

Work supported by the U.S. Department of Energy, Office of Science, Basic Energy Sciences, Materials Sciences and Engineering Division, and was performed at Oak Ridge National Laboratory's Center for Nanophase Materials Sciences (CNMS), a U.S. Department of Energy, Office of Science User Facility.

## Methods

### Sample Preparation

Atmospheric pressure chemical vapor deposition (AP-CVD) was used to grow graphene on Cu foil. The graphene was then coated in Poly(methyl methacrylate) (PMMA) for stabilization. For the heating experiments described here, a Protochips Fusion heater chip was used. To transfer the graphene onto the chip, the Cu foil was first etched away in an ammonium persulfate-deionized (DI) water bath. The remaining graphene/PMMA stack was scooped from the solution and rinsed with DI water to remove any residue from the ammonium persulfate, then lifted onto the Protochips heater chip. The chip was baked at 150 °C on a hot plate for ~15 min. to promote adherence. After cooling, the PMMA was dissolved away using acetone and finally rinsed with isopropyl alcohol. The sample was dried on the hot plate for a few minutes to ensure full evaporation of the solvents.

### Microscopy

The sample was examined in a Nion UltraSTEM U100 operated at 60 kV accelerating voltage. To clean the graphene *in situ,* the Protochips Fusion heating chip was ramped to 1200 °C at a rate of 1000 °C/ms following a protocol explored previously.[34]

EELS spectra were acquired with a convergence angle of 30 mrad and a collection angle of 48 mrad. Nominal beam current was in the range 60-70 pA.